\documentclass[10pt, conference]{IEEEtran}
\usepackage{braket}
\usepackage{cite}
\usepackage{graphicx}
\usepackage{color}
\usepackage[]{amsmath}
\usepackage{braket}
\graphicspath{{./images/}}
\begin{document}
\title{Sequential measurements on qubits by multiple observers: Joint best guess strategy}
\author{\IEEEauthorblockN{Dov Fields$^{1,2}$, \'Arp\'ad Varga$^{3}$, and J\'anos A. Bergou$^{1,2}$} \IEEEauthorblockA{$^{1}$Department of Physics and Astronomy, Hunter College of the City University of New York, \\ 695 Park Avenue, New York, NY, USA 10065 \\ $^{2}$Physics Program, Graduate Center of the City University of New York, \\
	365 Fifth Avenue, New York, NY 10016 \\ and \\ $^{3}$Institute of Physics, University of P\'ecs, \\ H-7624 P\'ecs, Ifj\'us\'ag \'utja 6, Hungary \\
	Emails: dovfields@gmail.com, vrg.arpad@gmail.com, jbergou@hunter.cuny.edu
}}
\maketitle
\begin{abstract} 
We study sequential state discrimination measurements performed on the same qubit by subsequent observers. Specifically, we focus on the case when the observers perform a kind of a minimum-error type state discriminating measurement where the goal of the observers is to maximize their joint probability of successfully guessing the state that the qubit was initially prepared in. We call this the joint best guess strategy. In this scheme, Alice prepares a qubit in one of two possible states. The qubit is first sent to Bob, who measures it, and then on to Charlie, and so on to altogether N consecutive receivers who all perform measurements on it. The goal for all observers is to determine which state Alice sent. In the joint best guess strategy, every time a system is received the observer is required to make a guess, aided by the measurement, about its state. The price to pay for this requirement is that errors must be permitted, the guess can be correct or in error. There is a nonzero probability for all the receivers to successfully identify the initially prepared state, and we maximize this joint probability of success. This work is a step toward developing a theory of nondestructive sequential quantum measurements and could be useful in multiparty quantum communication schemes based on communicating with single qubits, particularly in schemes employing continuous variable states. It also represents a case where subsequent observers can probabilistically and optimally get around both the collapse postulate and the no-broadcasting theorem.
\end{abstract}
\section{Overview}

Quantum state discrimination is an essential topic in Quantum Information theory with applications including, but not limited to, the secure distribution of information \cite{bennet} and the designing of experimental measurement schemes \cite{LBR}. While the laws of quantum mechanics do not allow perfect discrimination between two arbitrary nonorthogonal quantum states, these states can still be resolved probabilistically. There are a variety of possible criteria for quantum state discrimination. For recent reviews of state discrimination, see \cite{barnettrev,bergourev,baerev}. In this paper, we will focus on one particular discrimination procedure, minimum error state discrimination (ME). The ME strategy was first introduced in \cite{holevo,helstrom,yuen} as a protocol between two parties, Alice and Bob, involving two states (pure or mixed). In this strategy, Alice sends one of the two predetermined states to Bob according to a predetermined probability, and Bob's task is to perform a measurement with two possible outcomes. By correlating the outcomes of his measurement with the states that Alice sent, Bob can guess which state was sent by Alice at the cost of sometimes erroneously identifying the state sent. The goal of the ME strategy is for Bob to optimize his choice of measurement in a way that minimizes the average probability that he erroneously identifies the state sent by Alice. Further work on minimum error discrimination has focused on deriving the necessary and sufficient conditions for ME discrimination \cite{Barnett2009} and utilizing these conditions to understand the ME discrimination problem between more than two states \cite{Bae2013}.
\par
One recent development in the field of state discrimination is the sequential discrimination of states between multiple observers. The first extension of the standard unambiguous state discrimination (USD) to its sequential counterpart was proposed in Ref. \cite{BFH}. The work of this paper, which was complemented by Pang \emph{et al}., \cite{Pang2013}, focused on optimizing the sequential unambiguous state discrimination problem (SUD) for the case of equal prior probabilities. These works were further extended from qubits to qudits \cite{mimih2017} and led to investigations of the roles of quantum information \cite{rapcan} and quantum correlations and discord \cite{Zhang} in sequential measurements.
\par
In a recent paper \cite{Fields2020}, we completed the analysis of the SUD problem by generalizing the optimal solution to arbitrary prior probabilities. The present paper serves as a counterpoint to that work, in that we extend the ME strategy to its sequential equivalent for arbitrary priors. The sequential extension of the ME strategy is the joint best guess (JBG) strategy. The JBG protocol involves multiple parties, Alice, Bob, Charlie, etc. Alice sends Bob one of two predetermined states, and Bob chooses a Positive Operator Valued Measurement (POVM) such that he can have some probability of successfully determining the state sent by Alice at the cost of having some probability of error. Bob can design his POVM in such a way that when he sends his post-measurement states to Charlie, Charlie is able to then discriminate between these post-measurement states. This procedure allows Charlie an opportunity to also learn about the state sent by Alice. Additionally, this process can be repeated an arbitrary number of times so that there is a chain of N receivers who each, in sequence, receives the post-measurement states from their predecessor and can discriminate between them. The goal of this strategy is for all the receivers to optimize their joint probability of success. 

The goal of this paper is to consider the JBG strategy for N receivers for arbitrary priors. For simplicity, this analysis is restricted to only allowing pure post-measurement states so that each participant only needs to discriminate between two pure states. In the first section of this paper, we show how one can set up Bob's POVM so that Charlie can gain information from Bob's post-measurement states. Afterwards, we show how this can be extended to the problem of having N receivers, and derive the optimization problem and the related constraints for this setup. In the second part of this paper, we use basic optimization techniques to show that the general problem can be reduced to a simplified problem. In the last section of the paper, we derive an analytic solution for equal priors for the N-receiver problem and describe the numeric solution for arbitrary priors. We also consider the regions where these solutions are valid. Finally, we briefly consider an alternate strategy where, instead of maximizing the joint probability of success, each participant optimized their individual probability of succeeding.

\section{Setting up the POVM for Minimum Error}
\subsection{Two receivers}
For the problem of JBG discrimination between two non-orthogonal pure states, Alice sends the states $\ket{\psi_{1}}$ or $\ket{\psi_{2}}$ with probability $\eta_{1}$ and $\eta_{2}$, respectively. These states then get passed sequentially through a number of receivers, each of whom performs their own POVM on the states they receive from their predecessor, and then passes their post-measurement states along to the next link in the chain. The goal in this problem is to maximize the average probability that all of them succeed. To see how this sequence can be set up, it helps to start with the case of only two receivers, Bob and Charlie. In this situation, Bob's POVM takes the following form:
\begin{eqnarray}
	\sum_{i=1}^{2}\Pi_{bi} &=& I \label{completeness} \\
	\Pi_{bi} &\ge& 0 \ \ \  \mbox{for} \ \ \ i=1,2 \ .
\label{positivity}
\end{eqnarray}
Then $\braket{\psi_{i}|\Pi_{bi}|\psi_{i}} = p_{bi}$ is Bob's probability of correctly determining that state $i$ was sent and $\braket{\psi_{i}|\Pi_{j}|\psi_{i}} = r_{bi}$ for $i \neq j$ is Bob's probability of making an erroneous identification.  Because the POVM elements, $\Pi_{bi}$, are positive operators, we can write them in the form $\Pi_{bi} = B_{i}^{\dag}B_{i}$. The detection operators, $B_{i}$ determine the effect of Bob's POVM on the input states:
\begin{eqnarray}
	B_{1} &=&  \beta_{11}\ket{v_{11}}\bra{\psi_{2}^{\perp}} + \beta_{12}\ket{v_{12}}\bra{\psi_{1}^{\perp}}, \\
	B_{2} &=& \beta_{21}\ket{v_{21}}\bra{\psi_{2}^{\perp}} + \beta_{22}\ket{v_{22}}\bra{\psi_{1}^{\perp}}.
	\label{MEPOVMevolution}
\end{eqnarray}
Here $\ket{\psi_{i}^{\perp}}$ is the state orthogonal to $\ket{\psi_{i}}$, $\braket{\psi_{i}^{\perp}|\psi_{i}} = 0$.

While the problem can be completely determined from these conditions, it is convenient to represent Bob's POVM through the Neumark representation \cite{neumark,bergoubook}:

\begin{eqnarray}
	U_{b}\ket{\psi_{1}}\ket{0} &=& \sqrt{p_{1b}}\ket{v_{11}}\ket{1} + \sqrt{r_{1b}}\ket{v_{12}}\ket{2} 
	\label{BobNumarka} \\
	U_{b}\ket{\psi_{2}}\ket{0} &=& \sqrt{r_{2b}}\ket{v_{21}}\ket{1} + \sqrt{p_{2b}}\ket{v_{22}}\ket{2}
	\label{BobNumarkb}
\end{eqnarray}
Bob's measurement consists of entangling his qubit with an ancilla, $\ket{0}$ being a generic initial state of the ancilla, and then measuring the ancilla state. If he measures the ancilla to be in the state $\ket{i}$, corresponding to the $\Pi_{i}$ detector, Bob concludes that the state sent was $\ket{\psi_{i}}$ ($i=1,2$). Bob's probability of being correct given that state $\ket{\psi_{i}}$ was sent is $p_{ib}$. From unitarity $p_{ib} + r_{ib} = 1$ follows (again $i=1,2$). 
Taking the inner product of Eqs. \eqref{BobNumarka} and \eqref{BobNumarkb}, and making use of unitarity, one can derive the following constraint for Bob's probabilities:
\begin{eqnarray}
\braket{\psi_{2}|\psi_{1}} = \sqrt{p_{b1}r_{b2}}\braket{v_{11}|v_{12}} + \sqrt{p_{b2}r_{b1}}\braket{v_{21}|v_{22}}.
\label{constraintB}
\end{eqnarray}

After Bob's measurement, the qubit is in one of two mixed states, $\rho_{i} = p_{bi}\ket{v_{ii}}\bra{v_{ii}} + r_{bi}\ket{v_{ij}}\bra{v_{ij}}$ ($i=1,2$, $i \ne j$), depending on what state Alice sent. These states can then be discriminated by Charlie. If Bob chooses his POVM such that $\ket{v_{11}} = \ket{v_{12}} = \ket{v_{21}} = \ket{v_{22}}$ then Bob performs an optimal ME discrimination and leaves no possibility for Charlie to perform any type of discrimination. Alternatively, Bob can choose his POVM such that $\ket{v_{11}} = \ket{v_{12}}$ and $\ket{v_{22}} = \ket{v_{21}}$ ensuring that his output to Charlie is always a pure state. In this case, Bob has denied Charlie any possibility to learn the outcome of Bob's measurement, while still allowing Charlie to have a chance to guess the state initially sent by Alice. While Bob can also choose to send Charlie a set of mixed states for Charlie to discriminate, this paper will focus on the case where Charlie only has to discriminate between pure states.

After Bob's measurement, Charlie can perform a ME discrimination on the states resulting from Bob's POVM. In order for Charlie's measurement to be optimal, his POVM, $\Pi_{c}$, must be such that there is only one output state to his measurement. The remaining problem lies in choosing $\Pi_{b}$ and $\Pi_{c}$ such that the joint probability of both Bob and Charlie successfully identifying the state sent by Alice is optimized. Formally, one can state the problem as follows. Find the maximum of:

\begin{eqnarray}
	P_{ss} =  \sum_{i=1}^{2} \eta_{i}\braket{\psi_{bi}|\Pi_{bi}|\psi_{bi}}\braket{v_{i}|\Pi_{ci}|v_{i}} ,
\end{eqnarray}
subject to Eqs. \eqref{completeness} and \eqref{positivity} and their analogs for Charlie:
\begin{eqnarray}
	\sum_{i}^{N}\Pi_{ci} &=& I \label{completenessC}\\
	\Pi_{ci} &\ge& 0 \ \ \  \mbox{for} \ \ \ i=1,2 \ .
	\label{positivityC}
\end{eqnarray}
In this setup, Bob chooses a POVM such that when Alice sends $\ket{\psi_{i}}$ the output state is $\ket{v_{i}}$. This problem can be equivalently formulated in the following way. Find the maximum of:

\begin{eqnarray}
	P_{ss} &=&  \sum_{i=1}^{2} \eta_{i}p_{bi}p_{ci} ,
\end{eqnarray}
subject to:
\begin{eqnarray}
	\frac{s}{t} &=&  \sqrt{p_{b1}\left( 1 - p_{b2} \right)} + \sqrt{p_{b2}\left( 1 - p_{b1} \right)} \label{SMEbobconst}\\
	t &=& \sqrt{p_{c1}\left( 1 - p_{c2} \right)} + \sqrt{p_{c2}\left( 1 - p_{c1} \right)} \label{SMEcharlieconst}
	\label{Sequential Me, algebraic formalism}
\end{eqnarray}
where $p_{bi}$ and $p_{ci}$ are Bob and Charlie's probabilities of correctly identifying that Alice sent state $\ket{\psi_{i}}$, and where $s \equiv \braket{\psi_{1}|\psi_{2}}$ and $t \equiv \braket{v_{1}|v_{2}}$. Since there are only two states being discriminated, we can assume, without any loss of generality, that the overlaps between the states are real.

\subsection{N receivers}
From this example, it is clear how to extend the problem should to $N$ sequential receivers. Instead of choosing the optimal measurement, Charlie also needs to set up his POVM so that the post measurement states from his measurement are not identical. This is true for the first $N-1$ receivers, while the last receiver in the chain performs the optimal ME measurement on the states received from the previous observer. Assuming that the post-measurement states are restricted to pure states, the problem can then be formulated as follows. Find the maximum of the weighted Joint Best guess,
\begin{equation}
	P^{N}_{JBG} = \eta_{1}\prod^{N}_{n = 1}p_{1n} + \eta_{2}\prod^{N}_{n = 1}p_{2n},   
	\label{NJointSuccess}
\end{equation}
subject to the constraints
\begin{eqnarray*}
	\frac{t_{n}}{t_{n + 1}} &=& \sqrt{p_{1n}\left( 1 - p_{2n} \right)} + \sqrt{p_{2n}\left( 1 - p_{1n} \right)} \ \ \ n \le N-1 \\
	t_{N} &=& \sqrt{p_{1N}\left( 1 - p_{2N} \right)} + \sqrt{p_{2N}\left( 1 - p_{1N} \right)} .
	\label{NConstraints}
\end{eqnarray*}
Here $p_{in}$ is the $N^{th}$ receiver's probability of succeeding given that Alice sent state $\ket{\psi_{i}}$, $t_{1} \equiv \braket{\psi_{1}|\psi_{2}}$, and $t_{n}$ is the overlap of the post-measurement states received by the $N^{th}$ receiver.
\section{Simplifying the problem}
In order to optimize the JBG problem for arbitrary priors, it is important to realize that the problem can be simplified significantly. In order to show that this is possible, we need to use induction, starting from the case with only two receivers. Using Lagrange's principle, one can reformulate the SME problem for two receivers as follows:
\begin{eqnarray*}
	F &=& \eta_{1}p_{1b}p_{1c} + \eta_{2}p_{2b}p_{2c} \\
	&&+ \lambda_{1}\left( \frac{s}{t} - \left( \sqrt{p_{1b}\left( 1 - p_{2b} \right)} + \sqrt{p_{2b}\left( 1 - p_{1b} \right)} \right) \right) \\
	&&+ \lambda_{2}\left( t - \left( \sqrt{\left( p_{1c}\left( 1 - p_{2c} \right) \right)} + \sqrt{p_{2c}\left( 1 - p_{1c} \right)} \right) \right) 
	\label{LagrangeFunction}.
\end{eqnarray*}
Optimizing with respect to $p_{ib}$, $p_{ic}$, and $t$ gives five Lagrange conditions for the optimal solution:
\begin{eqnarray*}
	\eta_{i}p_{ic}\sqrt{p_{ib}\left(1 - p_{ib}\right)} = \frac{\lambda_{1}}{2}\left[\sqrt{\left(1 - p_{1b} \right)\left( 1 - p_{2b} \right)} - \sqrt{p_{1b}p_{2b}} \right], \\
	\eta_{i}p_{ib}\sqrt{p_{ic}\left(1 - p_{ic}\right)} = \frac{\lambda_{2}}{2}\left[\sqrt{\left( 1 - p_{1c} \right)\left( 1 - p_{2c} \right)} - \sqrt{p_{1c}p_{2c}} \right] 
	\label{LagrangeConditions}.
\end{eqnarray*}
\begin{eqnarray}
	\frac{s}{t^{2}} = \frac{\lambda_{2}}{\lambda_{1}}.
	\label{LagrangeSCondition}
\end{eqnarray}
Since the right hand sides of the first two sets of equations above do not depend on $i$, the left hand sides must also be equal, yielding
\begin{eqnarray}
	\eta_{1}p_{1c}\sqrt{p_{1b}\left( 1 - p_{1b} \right)} &=& \eta_{2}p_{2c}\sqrt{p_{2b}\left( 1 - p_{2b} \right)} \label{LagrangeEqualityone}\\
	\eta_{1}p_{1b}\sqrt{p_{1c}\left( 1 - p_{1c} \right)} &=& \eta_{2}p_{2b}\sqrt{p_{2c}\left( 1 - p_{2c} \right)} \label{LagrangeEqualitytwo}
	\label{LagrangeEqualities}
\end{eqnarray}
If we divide Eq. \eqref{LagrangeEqualitytwo} by Eq. \eqref{LagrangeEqualityone} and rearrange slightly, we get the relation: 
\begin{equation}
	\frac{\sqrt{p_{2b}\left( 1 - p_{1b} \right)}}{\sqrt{p_{1b}\left( 1 - p_{2b} \right)}} = \frac{\sqrt{p_{2c}\left( 1 - p_{1c} \right)}}{\sqrt{p_{1c}\left( 1 - p_{2c} \right)}}.
	\label{equalityrelation}
\end{equation}
By taking Eq. \eqref{SMEbobconst} and dividing by Eq. \eqref{SMEcharlieconst}, and using Eq. \eqref{equalityrelation}, we can derive that $\frac{s}{t^{2}} = \frac{\sqrt{p_{1b}\left( 1 - p_{2b} \right)}}{\sqrt{p_{1c}\left( 1 - p_{2c} \right)}}$. By equating this with $\frac{\lambda_{2}}{\lambda_{1}}$ using the final expression in Eq. \eqref{LagrangeConditions}, and dividing the second expression in Eq. \eqref{LagrangeConditions} by the first to get another formula for $\frac{\lambda_{2}}{\lambda_{1}}$, we can derive the following:
\begin{eqnarray*}
	\frac{\sqrt{\left( 1 - p_{1b} \right)\left( 1 - p_{2b} \right)}}{\sqrt{\left( 1 - p_{1c} \right)\left( 1 - p_{2c} \right)}}
	= \frac{\sqrt{\left( 1 - p_{1b} \right)\left( 1 - p_{2b} \right)} - \sqrt{p_{1b}p_{2b}}}{\sqrt{\left( 1 - p_{1c} \right)\left( 1 - p_{2c} \right)} - \sqrt{p_{1c}p_{2c}}}, \\
	 \sqrt{p_{2b}\left( 1 - p_{2c} \right)} = \frac{\sqrt{p_{1c}\left( 1 - p_{1b} \right)}}{\sqrt{p_{1b}\left( 1 - p_{1c} \right)}}\sqrt{p_{2c}\left( 1 - p_{2b} \right)}, \ \ \\
	p_{2c}\left( 1 - p_{2b} \right) = p_{2b}\left( 1 - p_{2c} \right) \Rightarrow p_{2c} = p_{2b}. \ \ \ \ \ \ \ \   
	\label{SMEsimpmath}
\end{eqnarray*}
In the second to last line, we used Eq. \eqref{equalityrelation} to derive the final line. Using this method, we can conclude that in the optimal solution $p_{1b} = p_{1c}$, $p_{2b} = p_{2c}$ and $t = \sqrt{s}$. In conclusion, the problem can be reduced to a simpler form of maximizing
\begin{equation}
	P_{ss} = \eta_{1}p_{1}^{2} + \eta_{2}p_{2}^{2} , 
	\label{SimplifiedEq}
\end{equation}
subject to the constraint:
\begin{equation}
	\sqrt{s} = \sqrt{p_{1}\left( 1 - p_{2} \right)} + \sqrt{p_{2}\left( 1 - p_{1} \right)} .
	\label{SimplifiedCon}
\end{equation}
In order to simplify the problem of $N$ receivers, we can use induction. Applying the Lagrange formalism to the JBG problem with $N$ receivers, the following optimization conditions can be derived in a straightforward manner:
{\belowdisplayskip=0pt
\begin{eqnarray*}
	\eta_{1}\prod^{N}_{j\ne i}p_{1j}\sqrt{p_{1i} \left( 1 - p_{1i} \right)}
	= \frac{\lambda_{i}}{2}\left[\sqrt{\left( 1 - p_{1i} \right)\left( 1 - p_{2i} \right)} - \sqrt{p_{1i}p_{2i}}  \right], \\ 
	\eta_{2}\prod^{N}_{j\ne i}p_{2j}\sqrt{p_{2i} \left( 1 - p_{2i} \right)} =\frac{\lambda_{i}}{2}\left[\sqrt{\left( 1 - p_{1i} \right)\left( 1 - p_{2i} \right)} - \sqrt{p_{1i}p_{2i}}  \right],
	\label{NLagrangeConstraints}
\end{eqnarray*}
\begin{eqnarray*}
	\quad \qquad \frac{t_{n}}{t_{n+1}^{2}} &=& \frac{\lambda_{n+1}}{\lambda_{n}t_{n+2}} \; n \le N-2, \\
	\quad \qquad \frac{t_{N-1}}{t_{N}^{2}} &=& \frac{\lambda_{N}}{\lambda_{N-1}}.
	\label{NLagrangeConstraintsCont}
\end{eqnarray*}
}
Focusing on the constraints for N and N-1, it can be shown that:
\begin{eqnarray*}
	&\eta_{1}\prod^{N-2}_{j}p_{1j}p_{1(N)}\sqrt{p_{1(N-1)}\left( 1 - p_{1(N-1)} \right)}  \\ 
	&= \eta_{2}\prod^{N-2}_{j}p_{2j}p_{2(N)}\sqrt{p_{2(N-1)}\left( 1 - p_{2(N-1)} \right)} , \\
	&\eta_{1}\prod^{N-2}_{j}p_{1j}p_{1(N-1)}\sqrt{p_{1(N)}\left( 1 - p_{1(N)} \right)} \\
	&= \eta_{2}\prod^{N-2}_{j}p_{2j}p_{2(N-1)}\sqrt{p_{2(N)}\left( 1 - p_{2(N)} \right)} , \\
	&\frac{\sqrt{p_{1(N}\left( 1 - p_{1(N - 1)} \right)}}{\sqrt{p_{1(N-1}\left( 1 - p_{1(N)} \right)}} = \frac{\sqrt{p_{2(N}\left( 1 - p_{2(N - 1)} \right)}}{\sqrt{p_{2(N-1}\left( 1 - p_{2(N)} \right)}}.
	\label{FurtherSubs}
\end{eqnarray*}
This final condition can be used to show that: 
\begin{equation*}
	\frac{t_{N-1}}{t_{N}^{2}} = \frac{\sqrt{p_{1(N-1)}\left( 1 - p_{2(N - 1)} \right)}}{\sqrt{p_{1(N)}\left( 1 - p_{2(N)} \right)}} = \frac{\lambda_{N}}{\lambda_{N-1}} .
	\label{NFinalEq}
\end{equation*}
In an identical fashion to the 2 receiver case, this can then be used to show that the optimal solution occurs for $p_{i(N-1)} = p_{i(N)}$ and $t_{N} = \sqrt{t_{n-1}}$. By substituting $t^{2}_{N}$ for $t_{N-1}$, the same procedure can be used N times to finally derive that for the optimal solution $t^{N}_{N} = t_{1} = s$. Using this result, we can show that this allows the JBG problem for N receivers to be simplified in the following form. Maximize
\begin{equation}
	P^{N}_{JBG} = \eta_{1}p_{1}^{N} + \eta_{2}p_{2}^{N},
	\label{Nsimplification}
\end{equation}
subject to the constraint
\begin{equation}
	s^{\frac{1}{N}} = \sqrt{p_{1}\left( 1 - p_{2} \right)} + \sqrt{p_{2}\left( 1 - p_{1} \right)} .
	\label{NConstraint}
\end{equation}
We note that this reduction is equivalent to maximizing the optimal probability of success for discriminating between two states with overlap $\braket{\tilde{\psi}_{1}|\tilde{\psi}_{2}} = s^{\frac{1}{N}}$, a total of $N$ times in a row.

\section{Optimizing for Arbitrary Priors}
While there is no simple analytic solution for optimizing the general JBG problem for arbitrary priors, the solution for equal priors, $\eta_{1} = \eta_{2} = \frac{1}{2}$, is known. The Lagrange constraints for equal priors can be satisfied when $p_{1} = p_{2} = \frac{1}{2}\left( 1 + \sqrt{1 - s^{\frac{2}{N}}} \right)$, yielding the optimal solution:
\begin{equation}
	P_{JBG,eq}^{N} = \left[\frac{1}{2}\left( 1 + \sqrt{1 - s^{\frac{2}{N}}}  \right)\right]^{N}  .
	\label{equalpriors}
\end{equation}

It is important to note that this analytic solution is only optimal for values of the overlap below a certain threshold. While the choice $p_{1} = p_{2}$ satisfies the constraints, it is not guaranteed to be a global maximum, only a local one. In order to understand the behavior of the full solution, we first describe the method to obtain the numerical solution. 
\begin{figure}[tbh]
	\centerline{\setlength{\unitlength}{1pt}
	\begin{picture}(280,240)(0,0)
		\put(10,205){\makebox(0,0){(a)}}
		\put(10,80){\makebox(0,0){(b)}}
		\put(10,185){\makebox(0,0){$P_{ss}$}}
		\put(10,65){\makebox(0,0){$p_{1}$}}
		\put(140,120){\makebox(0,0){$\eta_{1}$}}
		\put(140,0){\makebox(0,0){$\eta_{1}$}}
		\put(15,0){\includegraphics[width=\linewidth, height=120pt]{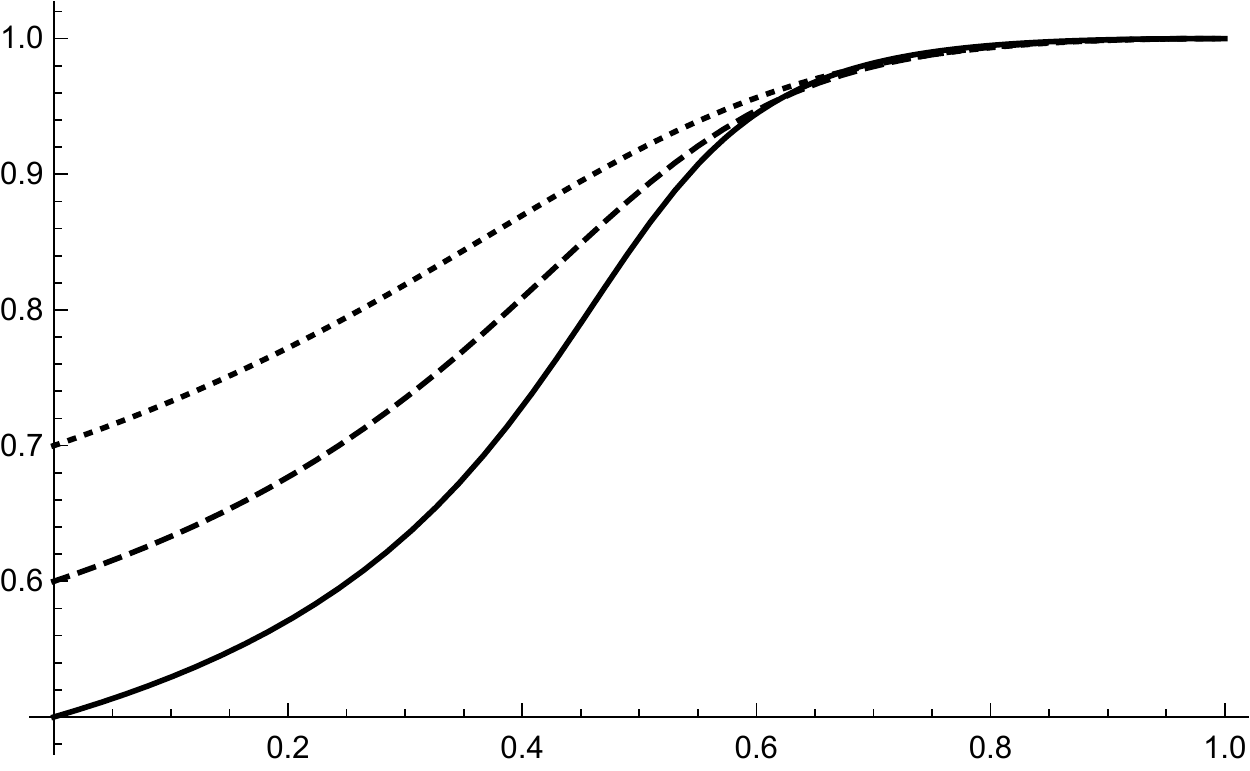}}
		\put(12.7,120){\includegraphics[width=\linewidth, height = 120pt]{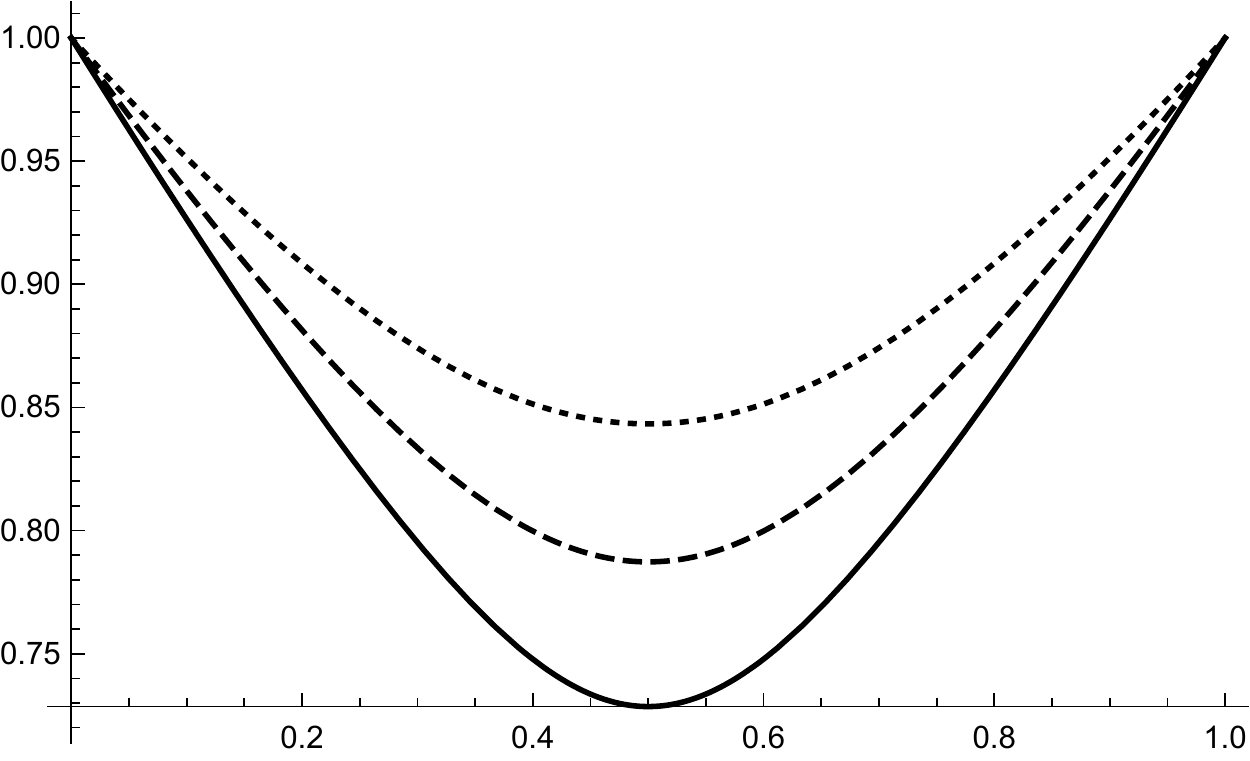}}
	\end{picture}}
	\caption{(a) Optimal joint probability of success, $P_{ss}$, and (b) Optimal success for the first state for two receivers, $_{1}$, versus the prior probability of sending the first state, $\eta_{1}$. Plots are calculated numerically for the values of the overlap: $s = 0.5$ (Solid), $s = 0.4$ (Dashed), and $s = 0.3$ (Dottted). Surprisingly, the term for $p_{1}$ does not go to zero when $\eta_{1} \rightarrow 0$. This feature is an artifact of constraint $\sqrt{s} = \sqrt{p_{1}\left( 1 - p_{2} \right)} + \sqrt{p_{2}\left( 1 - p_{1} \right)}$, which, for $p_{2} = 1$, requires $p_{1} = 1 - s$. However, $\eta_{1} p_{1} \rightarrow 0$ when $\eta_{1} \rightarrow 0$, as it should.}
	\label{fig:npsplots}
\end{figure}
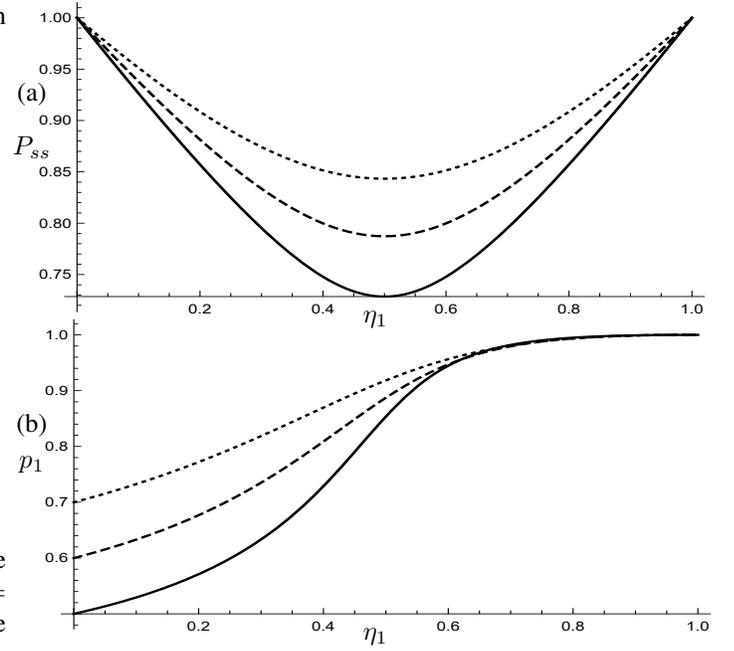
\begin{figure}[tbp]
	\centering
	\includegraphics[width = \linewidth]{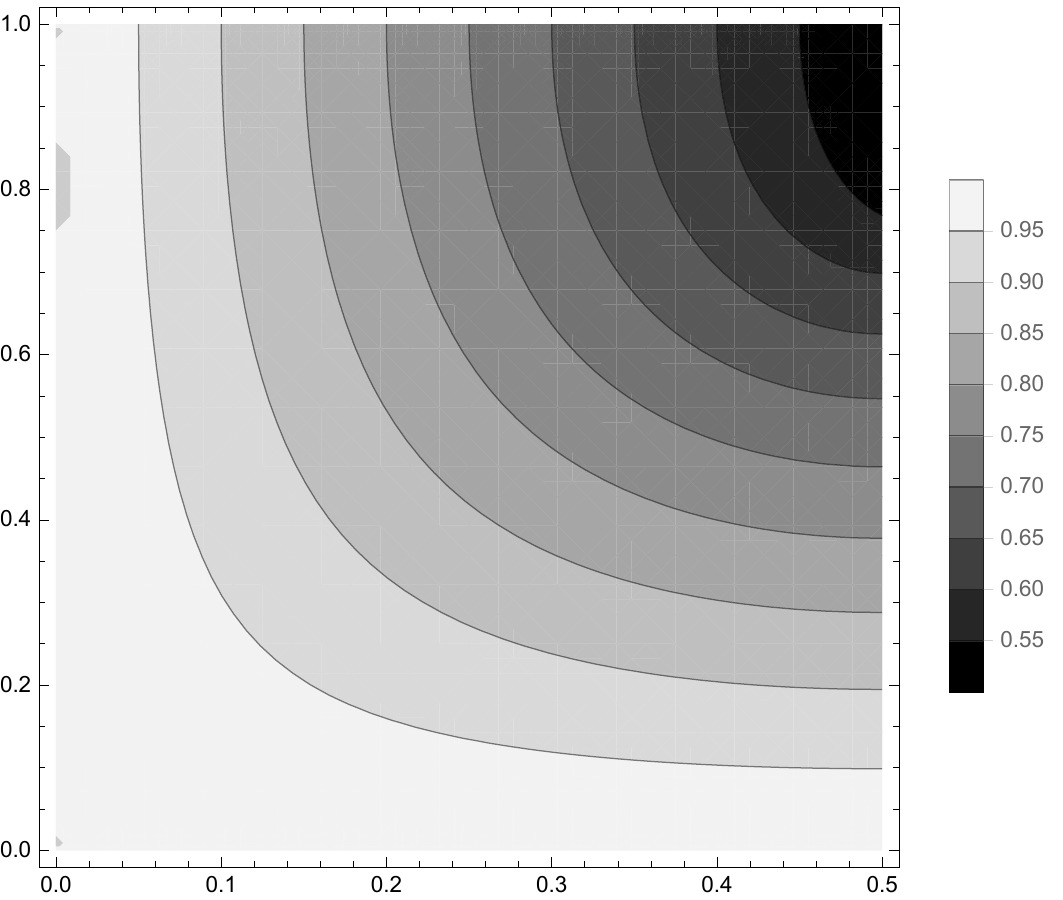}
	\put(-250,110){\makebox(0,0){$s$}}
	\put(-135,0){\makebox(0,0){$\eta_{1}$}}
	\put(-20,175){\makebox(0,0){$P_{ss}$}}
	\caption{A plot of the numerically derived optimal joint probability of success for two receivers as a function of the overlap and the prior probability }
	\label{fig:smecontour}
\end{figure}
\par
A complete solution can be derived numerically for both equal and arbitrary priors. This is shown for the case of two receivers in Figures \ref{fig:npsplots} and \ref{fig:smecontour}. Doing so requires rewriting the constraint in Eq. \eqref{NConstraint} for $p_{2}$. This can most easily be done by using the substitution $p_{i} = cos^{2} \left( \theta_{i} \right)$ and $s^{\frac{1}{n}} = \sin \left( \phi \right)$. Doing so rewrites the constraint as $\sin \left( \phi \right) = \sin \left( \theta_{1} \pm \theta_{2} \right)$, or $\phi = \theta_{1} \pm \theta_{2}$. Note that the two expressions for the constraint come from being able to choose both posive and negative square roots of $p_{i}$. While this gives four total possible expressions, there are only two distinct solutions. This gives two possible ways to express $p_{2}$:
\begin{eqnarray*}
	p_{2} &=& \cos^{2}\left( \phi + \theta_{1} \right) = \left(s^{\frac{1}{n}}\sqrt{1 - p_{1}} - \sqrt{p_{1}}\sqrt{1 - s^{\frac{2}{n}}}\right)^{2}, \nonumber\\
	&=& \cos^{2}\left( \phi - \theta_{1} \right) = \left(s^{\frac{1}{n}}\sqrt{1 - p_{1}} + \sqrt{p_{1}}\sqrt{1 - s^{\frac{2}{n}}}\right)^{2}.
	\label{p2sub}
\end{eqnarray*}
As the second of these two solutions for $p_{2}$ is always greater, it should therefore be the one chosen. By substituting this expression into the original joint success probability in Eq. \eqref{Nsimplification} and optimizing with respect to $\theta_{1}$, one derives the following expression:

\begin{eqnarray*}
	0 &=& \eta_{1}p_{1}^{n-1}\sqrt{p_{1}\left( 1 - p_{1} \right)} \\
	&+& \frac{\eta_{2}\left( s^{\frac{1}{n}}\sqrt{1 - p_{1}} + \sqrt{p_{1}}\sqrt{1 - s^{\frac{2}{n}}} \right)^{2n-1}}{ \left(\sqrt{\left( 1 - p_{1} \right)\left( 1 - s^{\frac{2}{n}} \right)} - s^{\frac{1}{n}}\sqrt{p_{1}}\right)^{-1}} .
	\label{smehomogeneouspoly}
\end{eqnarray*}

\begin{figure}[tb]
	\centering
	\includegraphics[width = \linewidth]{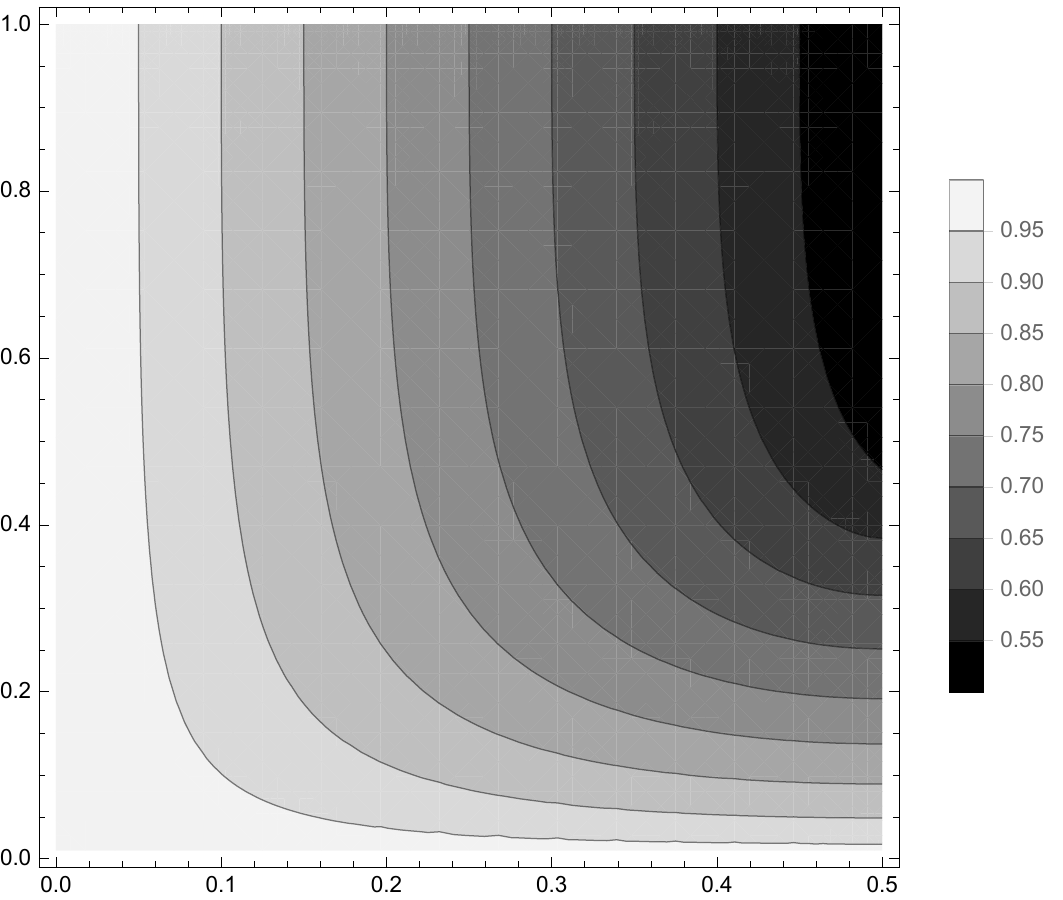}
	\put(-250,110){\makebox(0,0){$s$}}
	\put(-135,0){\makebox(0,0){$\eta_{1}$}}
	\put(-20,175){\makebox(0,0){$P_{ss}$}}
	\caption{Plot of the numerically derived optimal joint probability of success for three receivers as a function of the overlap and the prior probability.}
	\label{fig:threeplot}
\end{figure}

While optimizing with respect to $p_{1}$ gives an equation that is discontinuous at $p_{1} = 0$ and $p_{1} = 1$, optimizing with respect to $\theta_{1}$ avoids this problem. By solving this equation for $p_{1}$, one can obtain local optima of the joint success probability that can then be compared to the global optimum. For a visualization of the complete solution for two and three receivers, see Figures \ref{fig:smecontour} and \ref{fig:threeplot}. 

One possible concern for this numeric solution is that there might be boundary solutions for us to consider. The boundaries of this optimization problem are either ignoring one of the incoming states ($p_{i} = 0$) or prioritizing one of the incoming states ($p_{i} = 1$). However, the function is perfectly differentiable for all values $p_{i} \in \left[ 0,1 \right]$, ensuring that the boundary solution is accounted for by the internal point solution.
\par
\begin{figure}[tbh]
	\centering
	\includegraphics[width = \linewidth]{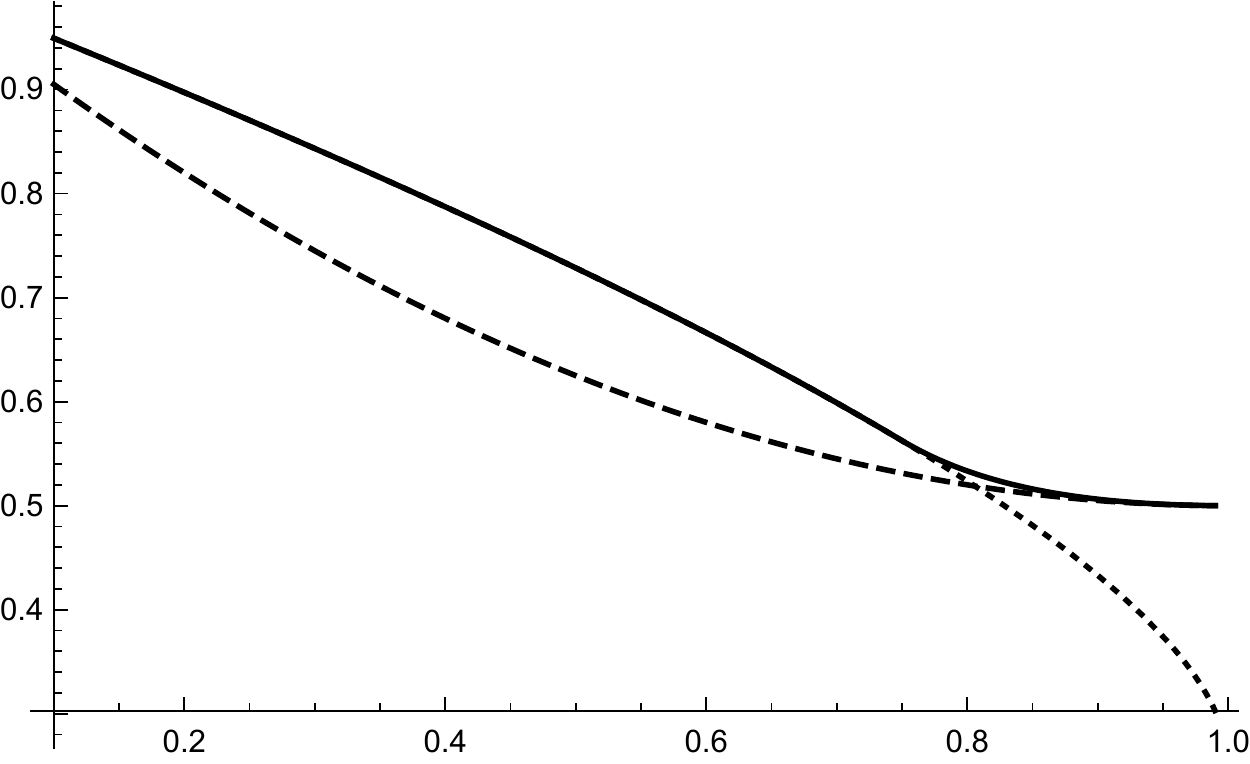}
	\put(-250,80){\makebox(0,0){$P_{ss}$}}
	\put(-120,0){\makebox(0,0){$s$}}
	\caption{A comparison of the joint probability of success as a function of the overlap for the numerically derived optimal solution (Solid), analytic solution (Dotted), and boundary solution (Dashed) for two receivers and equal priors.}
	\label{fig:boundaryplot}

\end{figure}

As noted earlier, while there is an analytic solution for the case of equal priors, it is only valid for values of the overlap $s<s_{b}$. When we numerically calculate $s_{b}$, we find that $s_{b} \approx 0.75$ for the case of two receivers and  $s_{b} \approx 0.42$ for the case of three receivers. In general, $s_{b}$ decreases as the number of receivers increases, decreasing the range of validity of the analytic solution. In Fig. \ref{fig:boundaryplot}, we compare the analytic solution, the numerically calculated optimal solution, and the boundary solution for the case of two receivers and equal priors.
\par
While the focus of this paper has been on the joint best guess strategy, there is a simpler approach that produces an analytic solution at a cost. Each member in the chain of receivers can simply optimize their own individual probability of succeeding ($P_{js}^{1}$) subject to the constraint given by Eq. \eqref{NConstraint}. From the standard Helstrom solution, doing so gives the probabilities of success
\begin{equation*}
	p_{i} =  \frac{1}{2}\left(1 + \frac{1 - 2(1 - \eta_{i})s^{\frac{2}{n}}}{\sqrt{1 - 4(\eta_{1}\eta_{2}s^{\frac{2}{n}})}} \right).
	\label{appxsol}
\end{equation*}

\begin{figure}[tb]
	\centerline{\setlength{\unitlength}{1pt}
	\includegraphics[width=\linewidth]{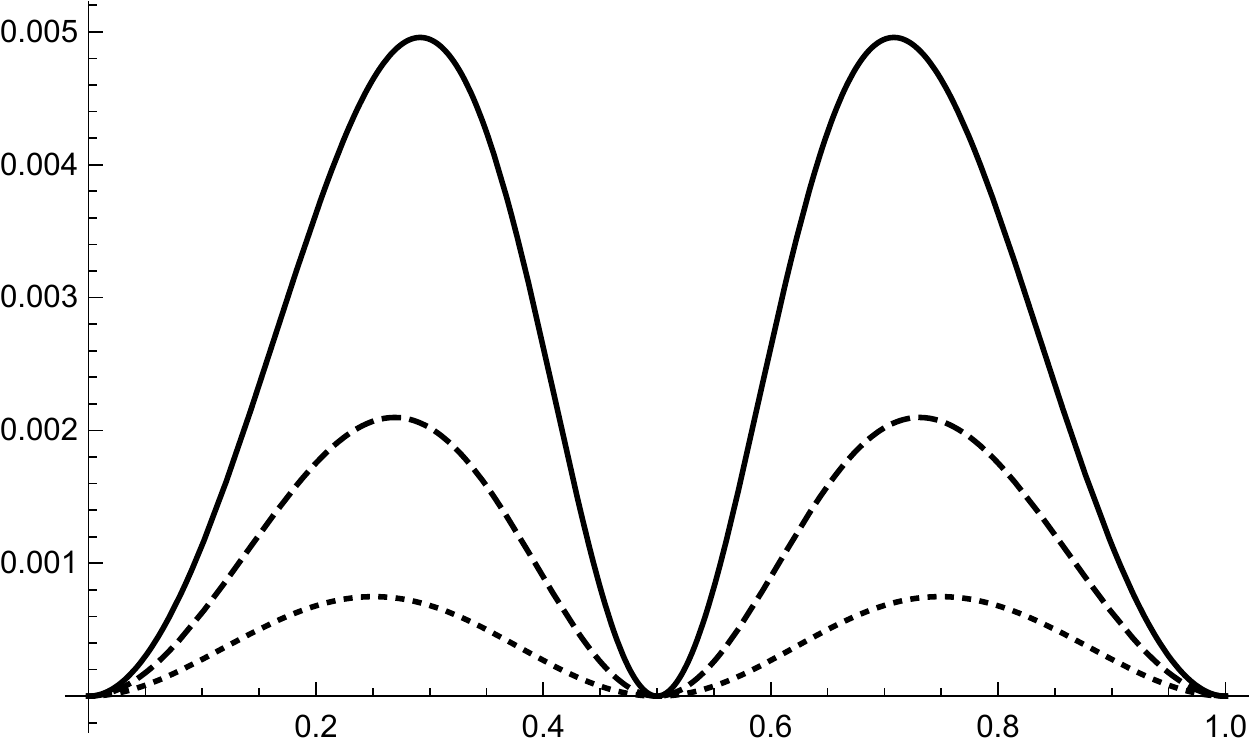}
	\put(-250,75){\makebox(0,0){$E$}}
	\put(-114,0){\makebox(0,0){$\eta_{1}$}}
}
	\caption{A plot of the $E = |P_{js,eq}^{2} - P^{2}_{js,appx}|$ as a function of prior probability $\eta$ for $s = 0.5$ (Solid), $s = 0.4$(Dashed), and $s = 0.3$(Dotted).}
	\label{fig:SMEappxeval}
\end{figure}

This solution ensures that each individual member maximizes their own average probability of succeeding at the cost of reducing the probability that they all simultaneously succeed. For the case of two receivers, the distinction between these two solutions is extremely small, as can be seen in Figure \ref{fig:SMEappxeval}. However, as the number of receivers increase, the cost of optimizing individual success to the joint success becomes more apparent.


\section{Conclusion}
In this paper, we contribute further analysis to the theory of sequential measurements. As a complement to the works in \cite{BFH} and \cite{Fields2020}, we present a complete analysis of the JBG discrimination problem for arbitrary prior probabilities. We start by explicitly showing how Bob can design a POVM that both lets him determine, with some probability, the state sent initially by Alice while also leaving enough information in the post-measurement states to pass on to Charlie who can then do the same. After deriving the constraints imposed by the chain of N-receivers, we show that in the optimal JBG strategy each participant in the chain discriminates between states with the same effective overlap. In other words, in the optimal strategy no participant is preferred and each succeeds and fails with the same probabilities. After deriving this critical result, we then show a full analysis of both the analytical and numerical solutions for the SME problem. Finally, we present an analytic solution that optimizes each participant's individual probability of success.
\par
This paper provides an alternative method of sequential communication to sequential unambiguous discrimination and serves a foundation for further research. The methods described in this paper could help answer open questions such as optimizing sequential measurement for mixed states and implementing other discrimination protocols sequentially.  

\par
Research was sponsored by the Army Research Laboratory and was accomplished under Cooperative Agreement Number W911NF-20-2-0097. The views and conclusions contained in this document are those of the authors and should not be interpreted as representing the official policies, either expressed or implied, of the Army Research Laboratory, or the U.S. Government. The U.S. Government is authorized to reproduce and distribute reprints for Government purposes notwithstanding any copyright notation herein.
\newpage
\bibliography{IEEEabrv,smebib}
\end{document}